\documentclass[twocolumn,showpacs,preprintnumbers,amsmath,amssymb,prl,superscriptaddress]{revtex4-1}

\usepackage{graphicx}
\usepackage{color}
\usepackage[colorlinks=true,plainpages=false,linkcolor=blue,urlcolor=blue,citecolor=blue,pdfpagemode=UseNone,pdfstartview=FitBH]{hyperref}

\begin{document}

\title{Crossover from Collective to Incoherent Spin Excitations in Superconducting Cuprates Probed by Detuned Resonant Inelastic X-ray Scattering}% Force line breaks with \\

\author{M. Minola}
\email[]{m.minola@fkf.mpg.de}
\affiliation{Max-Planck-Institut f\"{u}r Festk\"{o}rperforschung, Heisenbergstr. 1, 70569 Stuttgart, Germany}

\author{Y. Lu}
\affiliation{Max-Planck-Institut f\"{u}r Festk\"{o}rperforschung, Heisenbergstr. 1, 70569 Stuttgart, Germany}

\author{Y. Y. Peng}
\affiliation{CNISM, CNR-SPIN and Dipartimento di Fisica, Politecnico di Milano, 20133 Milano, Italy}

\author{G. Dellea}
\affiliation{CNISM, CNR-SPIN and Dipartimento di Fisica, Politecnico di Milano, 20133 Milano, Italy}

\author{H. Gretarsson}
\affiliation{Max-Planck-Institut f\"{u}r Festk\"{o}rperforschung, Heisenbergstr. 1, 70569 Stuttgart, Germany}

\author{M. W. Haverkort}
\affiliation{Max-Planck-Institut f\"{u}r Chemische Physik fester Stoffe, N\"{o}thnitzer Strasse 40, 01187 Dresden, Germany}
\affiliation{Institut f\"{u}r Theoretische Physik, Universit\"{a}t Heidelberg, Philosophenweg 19, 69120 Heidelberg, Germany}

\author{Y. Ding}
\affiliation{Beijing National Laboratory for Condensed Matter Physics and Institute of Physics, Chinese Academy of Sciences, Beijing 100190, China}

\author{X. Sun}
\affiliation{Beijing National Laboratory for Condensed Matter Physics and Institute of Physics, Chinese Academy of Sciences, Beijing 100190, China}

\author{X. J. Zhou}
\affiliation{Beijing National Laboratory for Condensed Matter Physics and Institute of Physics, Chinese Academy of Sciences, Beijing 100190, China}

\author{D. C. Peets}
\affiliation{Advanced Materials Laboratory, Fudan University, Shanghai 200438, China.}

\author{L. Chauviere}
\affiliation{Max-Planck-Institut f\"{u}r Festk\"{o}rperforschung, Heisenbergstr. 1, 70569 Stuttgart, Germany}
\affiliation{Department of Physics \& Astronomy, University of British Columbia, Vancouver British Columbia, V6T 1Z1, Canada}
\affiliation{Quantum Matter Institute, University of British Columbia, Vancouver, BC V6T 1Z4, Canada}

\author{P. Dosanjh}
\affiliation{Department of Physics \& Astronomy, University of British Columbia, Vancouver British Columbia, V6T 1Z1, Canada}
\affiliation{Quantum Matter Institute, University of British Columbia, Vancouver, BC V6T 1Z4, Canada}

\author{D. A. Bonn}
\affiliation{Department of Physics \& Astronomy, University of British Columbia, Vancouver British Columbia, V6T 1Z1, Canada}
\affiliation{Quantum Matter Institute, University of British Columbia, Vancouver, BC V6T 1Z4, Canada}

\author{R. Liang}
\affiliation{Department of Physics \& Astronomy, University of British Columbia, Vancouver British Columbia, V6T 1Z1, Canada}
\affiliation{Quantum Matter Institute, University of British Columbia, Vancouver, BC V6T 1Z4, Canada}

\author{A. Damascelli}
\affiliation{Department of Physics \& Astronomy, University of British Columbia, Vancouver British Columbia, V6T 1Z1, Canada}
\affiliation{Quantum Matter Institute, University of British Columbia, Vancouver, BC V6T 1Z4, Canada}

\author{M. Dantz}
\affiliation{Swiss Light Source, Paul Scherrer Institut, CH-5232 Villigen PSI, Switzerland}

\author{X. Lu}
\affiliation{Swiss Light Source, Paul Scherrer Institut, CH-5232 Villigen PSI, Switzerland}

\author{T. Schmitt}
\affiliation{Swiss Light Source, Paul Scherrer Institut, CH-5232 Villigen PSI, Switzerland}

\author{L. Braicovich}
\affiliation{CNISM, CNR-SPIN and Dipartimento di Fisica, Politecnico di Milano, 20133 Milano, Italy}

\author{G. Ghiringhelli}
\affiliation{CNISM, CNR-SPIN and Dipartimento di Fisica, Politecnico di Milano, 20133 Milano, Italy}

\author{B. Keimer}
\affiliation{Max-Planck-Institut f\"{u}r Festk\"{o}rperforschung, Heisenbergstr. 1, 70569 Stuttgart, Germany}

\author{M. Le Tacon}
\affiliation{Max-Planck-Institut f\"{u}r Festk\"{o}rperforschung, Heisenbergstr. 1, 70569 Stuttgart, Germany}
\affiliation{Institut f\"{u}r Festk\"{o}rperphysik, Karlsruher Institut f\"{u}r Technologie, Hermann-v.-Helmoltz-Platz 1, 76344 Eggenstein-Leopoldshafen, Germany}

\date{\today}

\begin{abstract}
Spin excitations in the overdoped high temperature superconductors Tl$_2$Ba$_2$CuO$_{6+\delta}$ and (Bi,Pb)$_2$(Sr,La)$_{2}$CuO$_{6+\delta}$ were investigated by resonant inelastic x-ray scattering (RIXS) as functions of doping and detuning of the incoming photon energy above the Cu-$L_3$ absorption peak. The RIXS spectra at optimal doping are dominated by a paramagnon feature with peak energy independent of photon energy, similar to prior results on underdoped cuprates.  Beyond optimal doping, the RIXS data indicate a sharp crossover to a regime with a strong
contribution from incoherent particle/hole excitations whose maximum shows a fluorescence-like shift upon detuning. The spectra of both compound families are closely similar, and their salient features are reproduced by exact-diagonalization calculations of the single-band Hubbard model on a finite cluster. The results are discussed in the light of recent transport experiments indicating a quantum phase transition near optimal doping.
\end{abstract}

%\pacs{75.20.Hr, 78.20.Ls, 78.70.Dm}

\maketitle

Recent transport experiments on a diverse set of materials including copper-based \cite{Ramshaw_Science2015,Badoux_Nature2016} and iron-based \cite{Kuo} high-temperature superconductors, layered chalcogenides \cite{LJLi}, and heavy-fermion intermetallics \cite{Gourgoux} have uncovered new evidence of quantum critical points (QCPs) in direct proximity to superconducting phases, and have thus galvanized research into the influence of quantum critical fluctuations on superconductivity. In particular, recent experiments on copper-oxide superconductors indicate a divergence of the electron mass \cite{Ramshaw_Science2015} and a Fermi surface reconstruction \cite{Badoux_Nature2016} near the doping level, $p_{opt}$, at which the superconducting transition temperature, $T_c$, is maximal. However, spectroscopic experiments monitoring potential mediators of superconductivity across the purported quantum critical points remain challenging. In the cuprates, neutron spectroscopy has revealed strongly temperature and doping dependent spin excitations \cite{Fujita_JPSJ2012} that have inspired magnetic pairing scenarios \cite{Scalapino_RMP2012,KeimerReview}.  However, most neutron experiments are limited to the underdoped regime, and a comprehensive investigation of the spin fluctuation spectrum across $p_{opt}$ has not been reported.

Resonant inelastic x-ray scattering (RIXS) has recently emerged as a powerful probe of magnetic and electron-hole excitations in transition metal compounds~\cite{Ament_RMP2011,BisogniNatComm2016}. RIXS has a lower energy resolution than neutron scattering, but its larger cross section and higher sensitivity to high-energy excitations have enabled the detection of dispersive spin excitations even in optimally \cite{LeTacon_NatPhys2011} and overdoped \cite{LeTacon_PRB2013,Dean_NatMat2013} cuprates. While these results lend general support to magnetic pairing theories, several important questions remain unanswered. First, RIXS experiments have not shown significant modifications of the spin fluctuation spectrum around $p_{opt}$, as one might have expected in the light of recent evidence of quantum criticality~\cite{Ramshaw_Science2015,Badoux_Nature2016}. In addition, the interpretation of the excitation profiles revealed by RIXS is still under debate. Whereas some model calculations indicate that these features arise from collective paramagnon modes induced by strong correlations \cite{Kung_PRB2015,Jia_PRX2016,WJLi}, others attribute them to incoherent, non-generic particle-hole excitations that reflect the band structure of specific cuprate compounds~\cite{Benjamin_PRL2014,Demler_PRB2016}. In the latter scenario, the matrix elements for photon-induced transitions between unoccupied and occupied states near the Fermi level (and hence the intensity maxima generated by dynamical nesting of the band structure) depend on the energy of the incoming photon. This results in a materials-dependent, fluorescence-like shift of the RIXS spectrum as the photon energy is varied. In contrast to this prediction, several recent experiments found that the maximum of the excitation profiles is independent of photon energy, as expected for Raman scattering from collective modes~\cite{Minola_PRL2015,Huang_SciRep2016} (see, however, Ref. \onlinecite{Guarise_NatCom2014}). So far, however, these ``detuning'' experiments have only been reported for $p \leq p_{opt}$, where the influence of electronic correlations is expected to be strong.

Here we address these open questions through systematic detuning measurements in Cu-$L_3$ edge RIXS on two different overdoped cuprate systems, (Bi,Pb)$_2$(Sr,La)$_{2}$CuO$_{6+\delta}$ (Bi2201) and Tl$_2$Ba$_2$CuO$_{6+\delta}$ (Tl2201). Combined with prior data on underdoped YBa$_2$Cu$_3$O$_{6+x}$ (YBCO), \cite{Minola_PRL2015} these measurements cover a wide range of doping with sufficiently high density to assess $p$-dependent modifications of the spin excitation spectra around optimal doping. Independent of the specific cuprate family, we find that the detuning dependence of the RIXS spectra changes rapidly from Raman-like to fluorescence-like as the doping level is tuned across $p_{opt}$, indicating a crossover from a collective response of the spin system to single-particle spin-flip excitations. This behavior is in qualitative accord with exact-diagonalization calculations of the two-dimensional Hubbard model. As a spectroscopic counterpart of recent transport experiments \cite{Ramshaw_Science2015,Badoux_Nature2016} , these results provide new perspectives for research on the interplay between quantum criticality and high-temperature superconductivity.

\begin{figure}%[t]
\includegraphics[width=0.98\columnwidth]{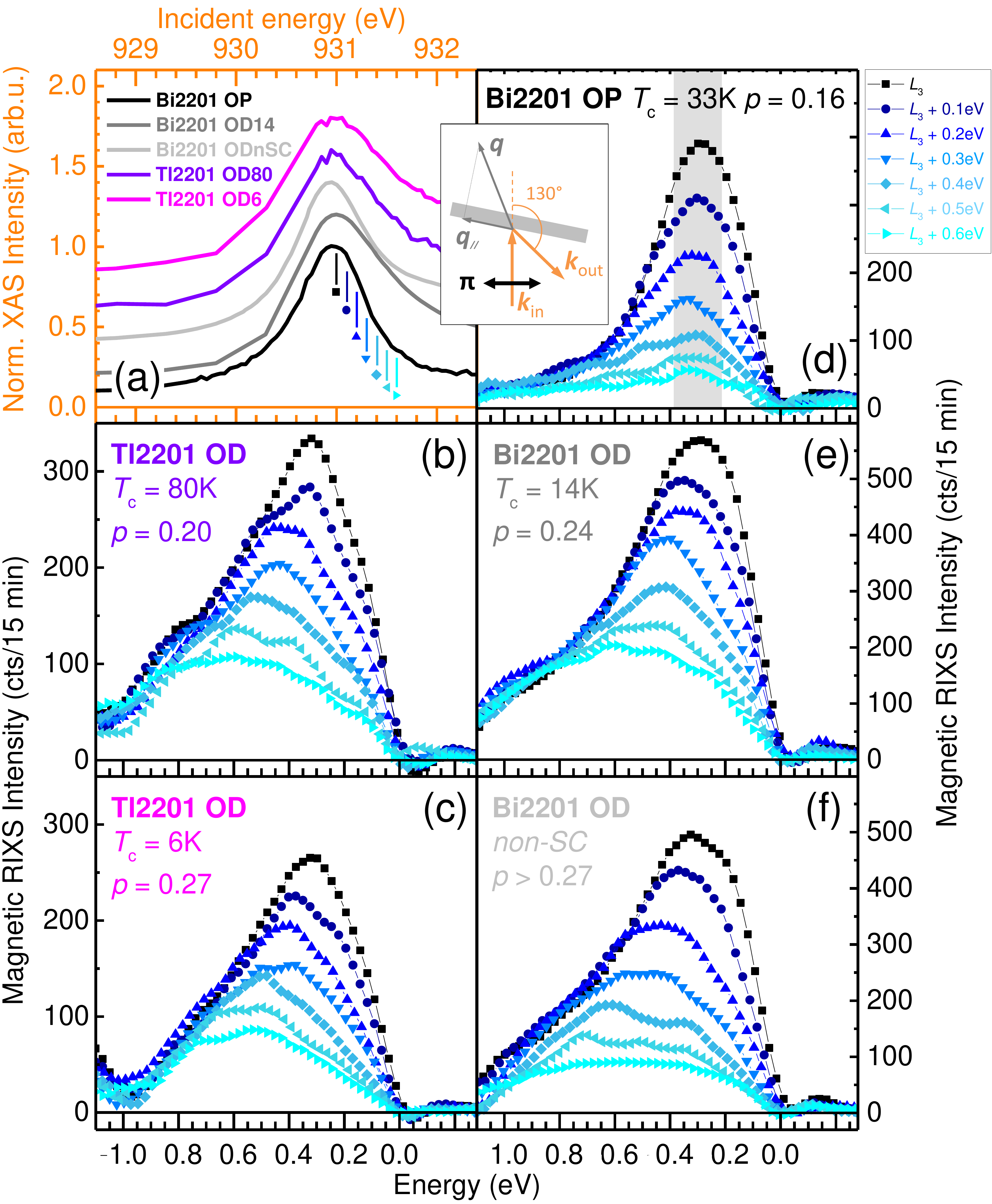}
\caption{\label{Fig1}(Color online) (a) Cu-$L_3$ edge x-ray absorption profiles measured near normal incidence in total electron yield for Bi2201 and Tl2201 samples (vertical offset for clarity). (b--f) Evolution of magnetic RIXS signal with $\pi$ polarization as function of detuning above the Cu-$L_3$ resonance for all samples investigated. Elastic line and tail from $dd$-excitations have been subtracted from the spectra. The inset shows the geometry of the RIXS experiments. The symbols and colors used for each detuning are reported also along the XAS profile in panel (a).}
%\vspace{-5mm}
\end{figure}

The RIXS measurements were carried out at the ADRESS beamline ~\cite{Strocov_JSR2010} of the Swiss Light Source (SLS) at the Paul Scherrer Institut, using the SAXES spectrometer~\cite{Ghiringhelli_RSI2006} in its high-throughput setting (180 meV resolution)~\cite{highthroughputSAXES}. We present RIXS data as a function of incident x-ray energy obtained on the following single crystals: optimally doped ($p$ = 0.16, $T_c$ = 33 K, henceforth referred to as OP33), lightly overdoped ($p$ = 0.24, $T_c = 14$  K, OD14), and heavily overdoped ($p>$ 0.27, non-superconducting, OD0) Bi2201, and lightly overdoped ($p$ = 0.20, $T_c$ = 80 K, OD80) and heavily overdoped ($p$ = 0.27, $T_c$ = 6 K, OD6) Tl2201. The hole content $p$ was obtained using Tallon's empirical formula~\cite{Tallon_PRB1995}. Details regarding the crystal growth and characterization are given in Refs.~\onlinecite{chineseBISCO} and~\onlinecite{Peets_JCrystGrowth2010}, for Bi2201 and Tl2201, respectively.

To determine the spin excitation spectra by RIXS, it is imperative to discriminate against charge excitations which exhibit a very different doping dependence.~\cite{Minola_PRL2015} Recent experiments with analysis of the scattered light polarization~\cite{Braicovich_RSI2014} have shown that $\pi$-polarization of the incident photons and grazing emission are optimal for the detection of electron spin-flip excitation in the $\pi\sigma^{\prime}$ scattering channel.~\cite{Minola_PRL2015} Here, $\pi$ ($\sigma$) denote polarization parallel (perpendicular) to the scattering plane. In the present work we have therefore used this geometry without any analysis of the scattered polarization. The data presented were obtained close to the antiferromagnetic zone boundary at momentum transfer ${\bf q}_{\parallel} = (0.37,0)$ in the CuO$_2$ planes (see inset of Fig.~\ref{Fig1} and Supplemental Material~\cite{supplemental}). When using $\sigma$ incident polarization in this geometry, the RIXS signal arises primarily from charge fluctuations~\cite{bimagnon} and does not contain single spin-flip excitations.
For all spectra presented we have fitted and subtracted the quasi-elastic peak and the background arising from the tail of electronic $dd$-excitations~\cite{Moretti_NJP2011}, in order to isolate the behavior of the low-energy inelastic spectral weight as function of the incident photon energy~\cite{supplemental}.

Figure~\ref{Fig1} shows the dependence of the magnetic RIXS signal of Bi2201 (panels a-c) and Tl2201 (panels e, f) as the incoming photon energy is detuned away from the Cu-$L_3$ resonance (931 eV, Fig.~\ref{Fig1}d). The key features of our study are clearly apparent in the raw data. In optimally doped Bi2201 (Fig. 1a), the excitation profile remains centered around  0.3 eV, the energy of the zone-boundary magnon in undoped antiferromagnetic cuprates~\cite{YYarXiv}, at all levels of detuning. Only a weak tail of the profile extends up to higher energies. This behavior reproduces prior RIXS experiments on other underdoped and optimally doped cuprates, which had revealed collective paramagnon excitations with a Raman-like detuning dependence~\cite{Minola_PRL2015}. In the overdoped samples, on the other hand, the high-energy tail strongly increases in intensity. The spectral weight of this tail is less strongly photon-energy dependent than that at 0.3 eV, so that both contributions become comparable in intensity at the highest levels of detuning. This results in a fluorescence-like shift of the overall spectral weight to higher energies upon detuning, but the lineshape of the RIXS spectra and their dependence on photon energy are more complex than predicted by independent-electron models. Moreover, the spectra of two distinctly different cuprate systems are closely similar, so that it is difficult to attribute the complex lineshape to specific features of the band structure of individual compounds. The data rather suggest that the spin excitation spectra are controlled by factors that are generic to the cuprates.

To compare the experimental data to a specific microscopic model, we have therefore numerically calculated the detuning-dependent RIXS spectra of the two-dimensional single-band Hubbard model, which has been widely used to describe the generic influence of correlations on the electronic structure of cuprates. The calculations were carried out by exact diagonalization (ED) on a 12-site Betts cluster with twisted boundary conditions within the many-body package {\sc Quanty},~\cite{ref1,ref2,ref3} which includes the effects of geometry and polarization. The hoppings were set to $t=0.4$~eV and $t^\prime=-0.2t$, and the on-site Coulomb repulsion $U=9t=3.6$~eV.  The core-valence interaction $U_c=7t$ and inverse core-hole life time $\Gamma=t$ were determined by fitting the calculated XAS spectra to the experimental ones~\cite{ref3,JapaneseDetuningTheory}. See the Supplementary Material~\cite{supplemental} for technical details and a discussion on the choice of parameters. The RIXS spectra are obtained with $\Gamma^\prime=0.005$~eV followed by a Gaussian broadening with FWHM=0.18~eV to reflect the instrumental resolution. Specifically, we calculated both the $\pi\sigma^{\prime}$ (spin-flip) and the $\pi\pi^{\prime}$ (charge) channel at ${\bf q}_{\parallel} = (1/3,0)$~\cite{supplemental}. As mentioned, in this geometry the $\pi\sigma^{\prime}$ scattering dominates the signal, but the $\pi\pi^{\prime}$ charge channel is known to have a residual contribution~\cite{Minola_PRL2015}. In order to evaluate this polarization leakage, we rescaled the bare $\pi\pi^{\prime}$ contribution by the geometrical factor $sin^2(20^{\circ})$, which accounts for the angle between the CuO$_2$ planes and the $\pi^{\prime}$ polarization vector of the emitted x-rays. The calculated RIXS spectra are then presented as sum of the $\pi\sigma^{\prime}$ scattering channel and $\pi\pi^{\prime}$ contribution~\cite{interference}.

\begin{figure}
\includegraphics[width=0.65\columnwidth]{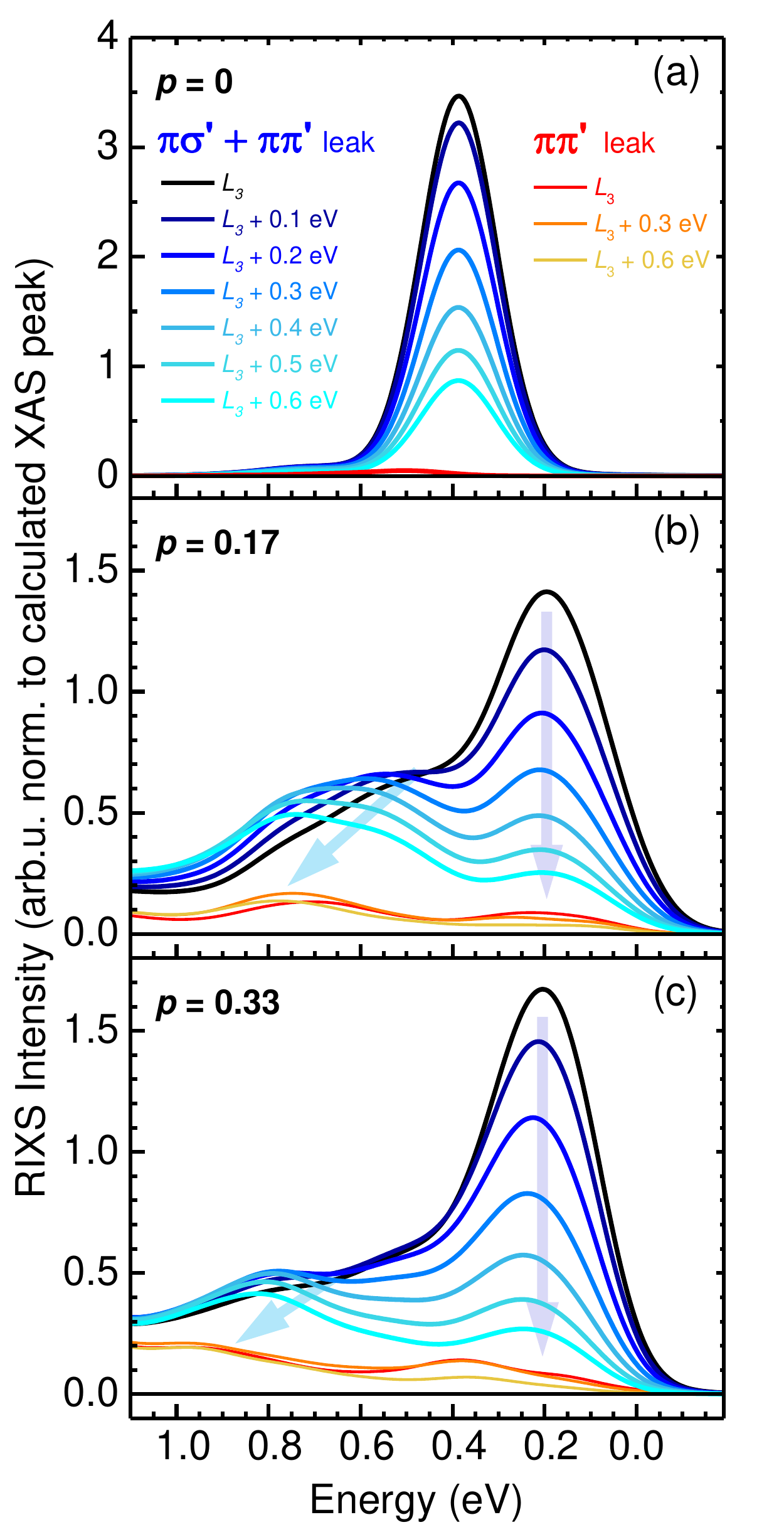}
 \caption{\label{Fig2}(Color online) Calculated RIXS spectra as sum of dominant $\pi\sigma^{\prime}$ channel and leakage from the $\pi\pi^{\prime}$ channel at $\bm{q}_\parallel=(1/3,0)$, for $p=$ 0, 0.17, and 0.33 as function of detuning. The orange lines depict the portion of spectra contributed by the $\pi\pi^{\prime}$ channel. The calculated spectra were broadened by a Gaussian with FWHM=0.18~eV to reflect the instrumental resolution.}
%\vspace{-5mm}
\end{figure}

Figure~\ref{Fig2} shows the calculated RIXS spectra for conditions matching those of the experimentally acquired data (Fig.~\ref{Fig1}), for $p= 0$, 0.17, and 0.33. The agreement between experimental and numerical data is remarkably good, considering that the computed results have no adjustable parameters apart from the overall scale factor. The calculations for the undoped case ($p= 0$) reproduce the experimental trend observed in undoped and underdoped YBCO~\cite{Minola_PRL2015}. The computed RIXS signals for $p=0.17$ and 0.33, on the other hand, are qualitatively similar to the experimental spectra for moderately ($p= 0.20$ and 0.24; Fig.~\ref{Fig1}c,d) and highly ($p= 0.27$ and $p>$ 0.27; Fig.~\ref{Fig1}e,f) overdoped cuprates, respectively. Specifically, the numerical data exhibit two components with a different detuning dependence: a paramagnon feature with a peak energy that is independent of detuning and cannot be described within a non-local band-structure approach~\cite{Benjamin_PRL2014}, and a tail that moves to higher energy with increasing detuning (arrows in Fig.~\ref{Fig2}b,c).

The experimental spectra cannot be unambiguously separated into two distinct components and can also be fitted with a single peak.\cite{supplemental} On the other hand, numerical and experimental data in the overdoped regime agree in the sense that the overall spectral weight shifts towards higher energy upon detuning in contrast to the underdoped and optimally doped samples where it remains centered at the paramagnon energy, independent of detuning. Furthermore, the signal broadens significantly upon detuning, again in qualitative agreement with the overall spectral weight in the numerical calculation. Future work will have to explore whether the discrepancies with the experimental data are consequences of the finite size of the numerically simulated lattice, or whether scattering channels not included in the single-band Hubbard model broaden the calculated spectra and wipe out the clear distinction between the two components~\cite{supplemental}.

The ED calculations have important implications for the possible interpretation of the RIXS data, even though they do not exclude other possible approaches discussed below. First, they demonstrate that spin excitations in a simple Hubbard model with realistic parameters generate a RIXS lineshape with a fluorescence-like component that increases strongly with increasing doping, in qualitative accord with the experimental observations.
Whereas the fluorescence-like feature resembles the prediction for the continuum of incoherent particle/hole excitations in independent-electron models, the Raman feature evolves smoothly into the magnon excitations for $p=0$~\cite{LeTacon_NatPhys2011,Minola_PRL2015} and can hence be attributed to correlation effects. Since the Hubbard model applies to all cuprates with only minor variations of the model parameters, the calculations also yield a natural explanation for the observation of nearly identical RIXS spectra in different cuprate families (Fig.~\ref{Fig1}).

The agreement of experimental (Fig.~\ref{Fig1}) and calculated (Fig.~\ref{Fig2}) spectra provides confidence in the capability of the numerical calculations to describe the salient features of the RIXS cross section. We have therefore used these calculations to estimate the contribution of charge excitations to the $\pi\pi^{\prime}$ scattering channel as function of detuning. Currently this quantity can not be measured experimentally, because the combination of polarimetry and detuning results in a prohibitively low count rate. The calculations show that the charge contribution in the experimental geometry is negligible for undoped and underdoped cuprates, whereas it accounts for up to $\sim 25$\% of the intensity at excitation energy $0.7$ eV for $p= 0.17$. This is in very good agreement with previous polarimetric measurements on slightly overdoped YBCO, which yielded an estimate of $\sim 20\%$ for the polarization leakage on resonance~\cite{Minola_PRL2015}. The maximal leakage occurs at the highest doping ($p= 0.33$), where about $40-50\%$ of the spectral weight arises from charge excitations. However, the polarization leakage depends only weakly on detuning, so that the intensity ratio of particle/hole-like and paramagnon-like spectral features (to be discussed below) is only weakly affected.

\begin{figure}
\includegraphics[width=0.9\columnwidth]{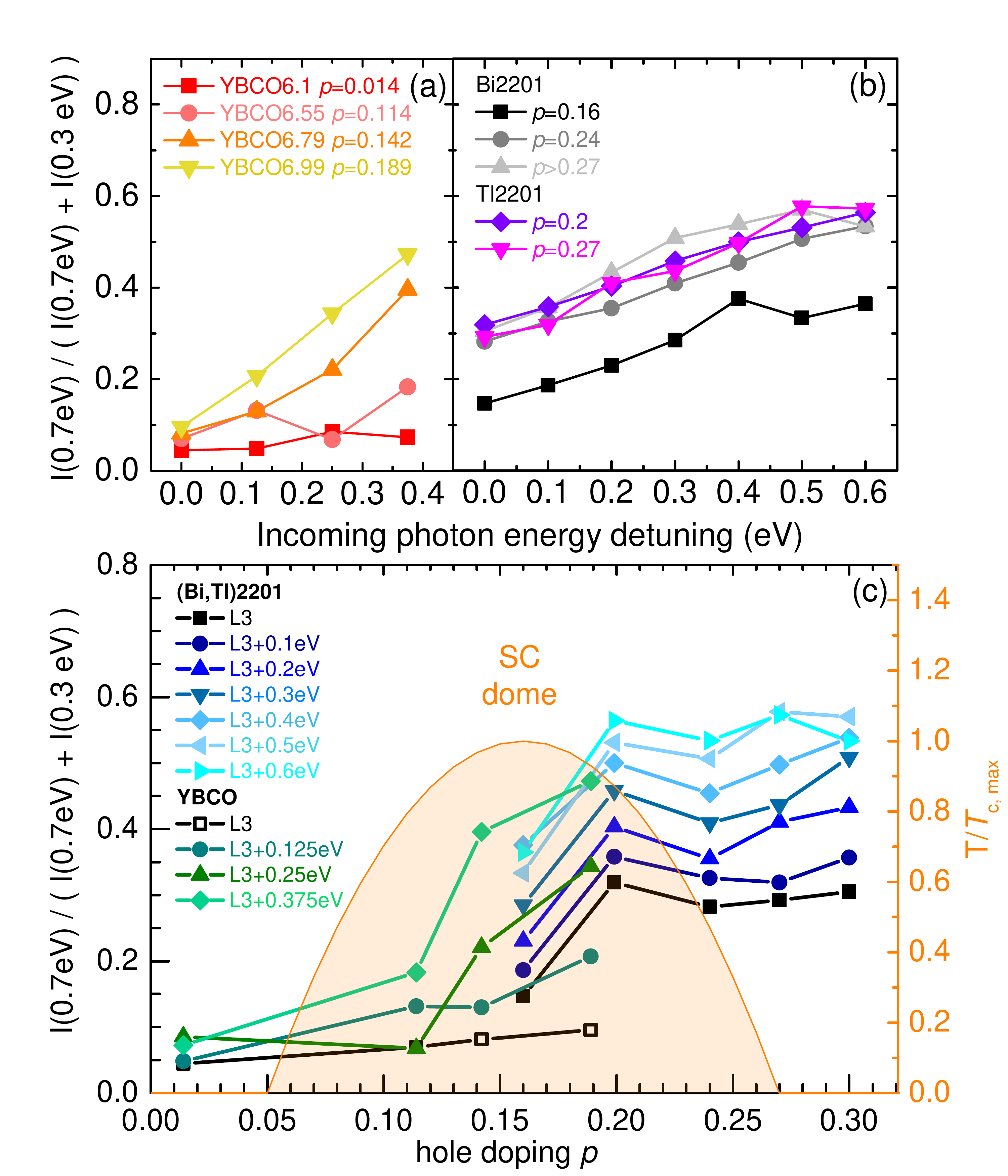}
 \caption{\label{Fig3}(Color online) Ratio of RIXS intensities at 0.7 and 0.3 eV excitation energy as function of detuning (a,b) and doping (c) for YBCO~\cite{Minola_PRL2015}, Bi2201, and Tl2201 (this work).}
%\vspace{-5mm}
\end{figure}

We now turn to the doping dependence of the spin excitations, using the intensity ratio at 0.3 and 0.7 eV as a simple, model-independent diagnostic of the integrity of the collective paramagnon mode. The ratio remains small as long as the spin-flip intensity is concentrated in a genuine collective mode with Raman-like detuning dependence at 0.3 V (Fig.~\ref{Fig1}a). It increases as soon as this mode loses intensity at the expense of more itinerant, continuum-like excitations – either by transferring its spectral weight to a separate high-energy component (as in Fig.~\ref{Fig2}) or by a detuning-induced shift of the 0.3 eV feature to higher energies~\cite{supplemental}. The detailed shape of the spectral features is not important for our conclusions. Figure~\ref{Fig3} displays the ratio $R=I(0.7 \, {\rm eV})/I(0.3 \, {\rm eV})$ of the measured intensities at these energies as function of detuning (top panels) and doping (bottom panel). To provide a comprehensive picture from the undoped antiferromagnet ($p =0$) to the highly overoped regime ($p \sim 0.3$), we have combined the current data set on overdoped Bi2201 and Tl2201 with prior data on undoped and underdoped YBCO taken under the same experimental conditions~\cite{Minola_PRL2015}. Both data sets overlap around optimal doping.

Figure~\ref{Fig3} shows that the detuning dependence of the spin excitation spectra changes rapidly in a narrow range around the crossover doping level $p_c \sim p_{opt} \sim 0.16$. For $p < p_c$, the detuning-independent, collective paramagnon mode at 0.3 eV dominates the spectra so that $R$ is small. As $p$ is tuned across $p_c$, the spectral weight of the high-energy component arising from the particle/hole continuum increases markedly. For $p > p_c$, $R$ saturates quickly, and the spectra evolve weakly with $p$ in the overdoped regime. In stark contrast with the pronounced doping dependence of the spin excitations, the charge excitations measured with $\sigma$ polarization always displays essentially the same fluorescent behavior across the entire phase diagram~\cite{supplemental}. Spin and charge excitations thus exhibit more closely parallel behavior in the overdoped regime, as expected for a Fermi liquid. However, a residual paramagnon component with Raman-like detuning dependence seems to persist even for highly overdoped samples, preventing the magnetic RIXS signal from having a fully fluorescent behavior ($R \approx 1$) upon detuning.

In summary, RIXS detuning experiments spanning the complete phase diagram of the cuprate superconductors have revealed the rapid rise of a spin excitation continuum close to optimal doping, on top of a paramagnon mode that persists throughout the phase diagram. This behavior is observed in multiple cuprate compounds \cite{differences}, suggesting that it is generic for the cuprates. The crossover from the collective spin dynamics characteristic of strongly correlated electrons to a regime more closely resembling the particle/hole continuum of a weakly correlated electron system is strikingly analogous to the results of recent experiments that indicate a major reconstruction of the electronic structure around optimal doping. Specifically, above $p\sim0.15$, ARPES has for instance reported a recovery of the nodal quasiparticle spectral weight $Z_N$ to the Fermi liquid expected value $2p/(p+1)$~\cite{Damascelli_NatPhys2010} and recent measurements of the Hall coefficient in high magnetic fields~\cite{Badoux_Nature2016} indicate a quantum phase transition associated with a change in carrier density $n$ from $n=p$ to $n=1+p$ for $p$ between 0.16 and 0.19, consistent with the crossover doping level $p_c$ found in our RIXS experiment. While the origin of this transition is still a matter of intense debate, it is tempting to associate the spin fluctuation continuum revealed by RIXS with the additional carriers found in the Hall effect measurements. It will be interesting to explore whether the relative loss of paramagnon spectral weight observed by RIXS weakens the spin fluctuation mediated pairing interaction enough to explain the decrease of the superconducting transition temperature in the overdoped regime. Based on the spectroscopic data we have presented, this question can now be addressed in a systematic fashion.\\

%%%%%%%%%%%%%%%%%%%%%%%%%%%%%%%%%%%%%
M.M. and Y.L. contributed equally to this work. %We acknowledge K. Wohlfeld for fruitful discussions.
The experiments were carried out at the ADRESS beamline using the SAXES instrument jointly built by the
Paul Scherrer Institut (Villigen, Switzerland), Politecnico di Milano (Italy) and Ecole
polytechnique f\'ed\'erale de Lausanne (Switzerland).
M.M. was supported by the Alexander von Humboldt Foundation.
YYP, GD and GG were supported by the PIK-POLARIX project of the Italian Ministry of Research (MIUR).
Part of this research has been funded by the Swiss National Science Foundation through the Sinergia network Mott Physics Beyond the Heisenberg (MPBH) model and and the D-A-CH program (SNSF Research Grant No. 200021L 141325). The research leading to these results has also received funding from the European Community's Seventh Framework Program (FP7/2007-2013) under Grant Agreement No. 290605 (COFUND: PSI-FELLOW).
The work at UBC was supported by the Max Planck-UBC Centre for Quantum Materials, the Killam, A. P. Sloan, A. von Humboldt, and NSERC's Steacie Memorial Fellowships (A.D.), the Canada Research Chairs Program (A.D.), NSERC, CFI and CIFAR Quantum Materials.
%%%%%%%%%%%%%%%%%%%%%%%%%%%%%%%%%%%%%%


\begin{thebibliography}{33}
\bibitem{bimagnon} With $\sigma$ incident polarization and grazing emission the RIXS signal arises mostly from charge fluctuations but there is also a contribution from multi-magnons with even parity, primarily bimagnons, that get progressively overdamped and rapidly negligible for increasing doping.
\bibitem{interference} The simple sum of the calculated $\pi\sigma^{\prime}$ and $\pi\pi^{\prime}$ scattering channels is an extremely good approximation
of the actual experimental RIXS signal since the interference effects between the two channels are negligible in the grazing emission geometry of this work.
\bibitem{broad} Since the RIXS profiles are much broader than the experimental resolution, the origin of the broadening must be intrinsic. This is no surprising per se, because the spin correlations may be affected by disorder and by coupling to phonons, etc., and the full band structure is more complex than the single-band Hubbard model assumed in the calculations. These details will have to be addressed in future theoretical work.
\bibitem{differences} Notice that YBCO, Bi2201 and Tl2201 have very different $T_c$, chemical disorder (higher in Bi2201), and phenomenology of competing phases such as charge order, which is ubiquitous in hole doped cuprates, but stronger in YBCO than in Bi2201.
\end{thebibliography}
\end{document}